\documentclass[twocolumn,pre,aps,showpacs,amsmath,amssymb,superscriptaddress]{revtex4-1}

\usepackage{hyperref}
\usepackage{amsmath,amssymb}
\usepackage{amsfonts,amsthm}
\usepackage{graphics}
\usepackage{graphicx}
\usepackage{dcolumn}
\usepackage{color}
\usepackage{bm}
\usepackage[normalem]{ulem}

\begin{document}

\title{Feedforward and feedback influences through distinct frequency bands between two spiking-neuron networks}

\author{Leonardo Dalla Porta}
\thanks{Equally contributed}
\affiliation{Systems Neuroscience, Institut d'Investigacions Biom\`ediques August Pi i Sunyer (IDIBAPS), Barcelona, Spain.}
\author{Daniel M. Castro}
\thanks{Equally contributed}
%Daniel must decide how to show his name
\affiliation{Departamento de F{\'\i}sica, Universidade Federal de Pernambuco (UFPE),
Recife, PE, Brazil.}
\author{Mauro Copelli}
\affiliation{Departamento de F{\'\i}sica, Universidade Federal de Pernambuco (UFPE),
Recife, PE, Brazil.}\author{Pedro V. Carelli}
\thanks{pedro.carelli@ufpe.br}
\affiliation{Departamento de F{\'\i}sica, Universidade Federal de Pernambuco (UFPE),
Recife, PE, Brazil.}
\author{Fernanda S. Matias}
\thanks{fernanda@fis.ufal.br}
\affiliation{Instituto de F\'{\i}sica, Universidade Federal de Alagoas (UFAL), Macei\'{o}, AL, Brazil.}

%%%%%%%%%%%%%%%%%%%%%%%%%%%%%%%%%%%%%%%
%%%%%%%%%%%%%%%%%%%%%%%%%%%%%%%%%%%%%%%
%%%%%%%%%%%%%%%%%%%%%%%%%%%%%%%%%%%%%%%
\begin{abstract}
Several studies with brain signals suggested that bottom-up and top-down influences are exerted through distinct frequency bands
among visual  cortical areas. It has been recently shown that theta and gamma rhythms subserve feedforward, whereas the feedback influence is dominated by the alpha-beta rhythm in primates.
A few theoretical models for reproducing these effects have been proposed so far.
Here we show that a simple but biophysically plausible two-network motif composed of spiking-neuron models and chemical synapses can exhibit feedforward and feedback influences through distinct frequency bands.
Differently from previous studies, this kind of model allows us to study directed influences not only at the population level, by using a proxy for the local field potential, but also at the cellular level, by using the neuronal spiking series. 
\end{abstract}
\maketitle

%%%%%%%%%%%%%%%%%%%%%%%%%%%%%%%%%%%%%%%
%%%%%%%%%%%%%%%%%%%%%%%%%%%%%%%%%%%%%%%
%%%%%%%%%%%%%%%%%%%%%%%%%%%%%%%%%%%%%%%
\section{Introduction}

Understanding the relationship between structural and functional connectivity in the brain is one of the greatest challenges of neuroscience. In other words, this means associating specific anatomical networks with different possible patterns of activity and especially with the information flow.
Regarding the hierarchical organization of cortical regions, several anatomical studies have shown that the structural connections from the primary sensory areas to higher-order areas (i.e. feedforward or bottom-up direction) are reciprocated by connections in the opposite direction (known as feedback or top-down connections) ~\cite{Felleman91,Markov14}. 
Furthermore, many cognitive phenomena including visual attention and perception have been related to both feedforward and feedback influences ~\cite{Posner80,Moran85}.

In visual cortical areas of primates, the hierarchy is reflected not only in their projection patterns along with different cortical layers (anatomical connectivity) but also concerning local rhythmic synchronization (functional connectivity). For example, feedforward projections typically originate from superficial layers, whereas feedback projections originate predominantly from infragranular layers. Synchronization in the gamma frequency band is strongest in superficial layers, whereas synchronization in the alpha-beta frequency band is strongest in infragranular layers~\cite{Buffalo11,Roberts13,Xing12}. Moreover, electrical stimulation of the cortical area V1 induces enhanced oscillatory activity in the cortical region V4 in the gamma-band, whereas the stimulation of V4 induces enhanced alpha-beta-band activity in V1~\cite{VanKerkoerle14}. Taken together, these results suggest that gamma might subserve feedforward and alpha-beta feedback information flow~\cite{Lee13,Fries05}
 
Two recent experimental studies with primates have corroborated these ideas. First, it has been shown through large-scale high-density electrocorticography and anatomical projection patterns in rhesus macaques that among 8 visual cortical areas gamma is systematically stronger in the feedforward direction and beta in the feedback direction~\cite{Bastos15}. 
Second, it has been reported that in human visual areas feedforward or feedback could be determined based on retrograde tracing data in homologous macaque visual areas. Moreover, by using magnetoecephalogram data and spectral Granger causality analysis to determine causal influences among cortical areas Michalareas et al.~\cite{Michalareas16} have shown that feedforward projections were predominant in the gamma band, whereas feedback projections were predominant in the alpha-beta band.

The mechanisms underlying these frequency-specific influences are still under investigation. A large-scale network using mean-field rate models of the Wilson-Cowan type have been employed to reproduce the feedforward and feedback influences through distinct frequency bands among 8 selected cortical areas of interest (V1, V2, V4, DP, 8m, 8l, TEO, and 7A)~\cite{Mejias16}.
Even though they reproduced the Granger causality patterns observed by Bastos et al.~\cite{Bastos15} among the mentioned regions, a firing rate model cannot be used to study the spiking time relations between neurons in different areas, which is a natural future step of investigation in experimental studies~\cite{Bastos15}. Furthermore, spiking neuron models allow us to investigate the effect of neuronal heterogeneity and homeostasis. In such direction, Lee et al.~\cite{Lee13,Lee15} have developed a  biophysically based model to show how top-down signals in the beta and gamma regimes can interact with a bottom-up gamma rhythm to provide regulation of signals between the cortical areas and among layers. However, in their studies, they were interested in reproducing in vitro observations between the primary auditory cortex and adjacent association cortex and did not reproduce the unidirectional gamma feedforward and unidirectional alpha-beta feedback verified in visual areas~\cite{Bastos15,Michalareas16}.
 
Here we show that two reciprocally connected cortical-like populations composed of randomly connected Izhikevich neurons can present Granger causality from 1 to 2 in the gamma band whereas the influence from 2 to 1 occurs in the alpha band. 
In Sec.~\ref{methods}, we describe the neuronal population model as well as the parameters that we use to change inter-areas coupling.
We also describe the spectral time series analysis that we use to characterize influences at the two different spatial scales: populational and neuronal levels.
In Sec.~\ref{results}, we report our results, showing that our motif can exhibit bottom-up and top-down influences in alpha ($\sim 10$~Hz) and gamma ($\sim 40$~Hz) bands. 
One of the advantages of a spiking-neuron network model is that one can explore information measures at the neuronal level such as the directional spike-train pairs associated with directional information (DI)~\cite{Campo19}.
In fact, we employ this method and show that the spiking neuron trains can give us complementary information about the direction of influence.
Concluding remarks and a brief discussion of the significance of our findings for neuroscience are presented in Sec.~\ref{conclusions}.

%%%%%%%%%%%%%%%%%%%%%%%%%%%%%%%%%%%%%%%
%%%%%%%%%%%%%%%%%%%%%%%%%%%%%%%%%%%%%%%
%%%%%%%%%%%%%%%%%%%%%%%%%%%%%%%%%%%%%%%
\section{Methods}
\label{methods}

\begin{figure}[!ht]%
\includegraphics[width=1\columnwidth,clip]{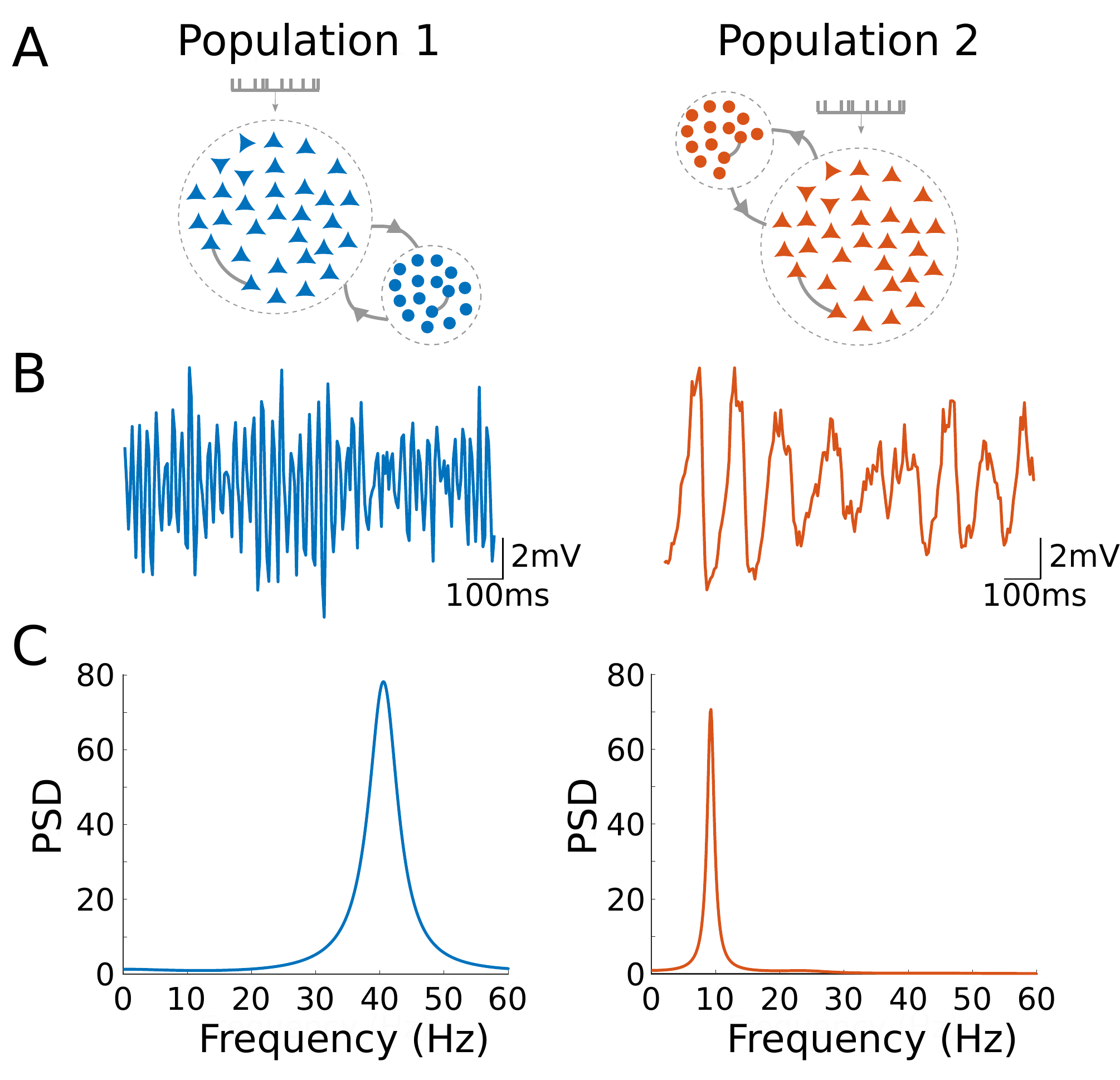}
\caption{\label{fig:uncoupled} Cortical motif circuits. (A) Schematic representation of two uncoupled cortical populations, $1$ (left, blue) and $2$ (right, orange), with excitatory and inhibitory neurons represented in triangles and circles, respectively. (B) Example of oscillatory activity for both populations, represented by the average membrane potential. (C) Power spectrum of the average membrane potential depicted in (B) (see Methods for details).}
\end{figure}%

\subsection{Modeling the spiking-neuron networks}

We modelled two neuronal populations following the ideas proposed in Ref~\cite{Matias14}. Both populations, namely $1$ and $2$, consist of $400$ excitatory and $100$ inhibitory neurons each. Each neuron is modeled by the Izhikevich model \cite{Izhikevich03}:
\begin{eqnarray}
  \frac{dv}{dt} &=& 0.04v^2+5v+140-u +\sum I_{syn} + I_{DC}, \label{dv/dt}
\end{eqnarray}
and 
\begin{eqnarray}
  \frac{du}{dt} &=& a(bv-u). \label{du/dt}
\end{eqnarray}
If $v\geq30$~mV, then $v$ is reset to $c$ and $u$ to $u+d$. $v$ and $u$ stand for the membrane potential and the membrane recovery variable (activation of K$^+$ and inactivation of Na$^+$ ionic currents), respectively. $a,b,c$ and $d$ are dimensionless parameters that account for the firing patterns heterogeneity which are randomly distributed accordingly to the neuron's nature. For excitatory neurons $a=0.02$, $b=0.20$, $c=-65+15\sigma^2$ and $d=8-6\sigma^2$, whereas for inhibitory neurons $a=0.02+0.08\sigma$, $b=0.25-0.05\sigma$, $c=-65$ and $d=2$, where $\sigma\in(0,1)$ is a random variable. Additionally, all neurons have $I_{DC}=0$ except for the excitatory neurons in population $1$, which are submitted to a constant current of $I_{DC}=25$~pA.

The synaptic transmissions are mediated by excitatory AMPA (A) and fast inhibition GABA$_A$ (G).
The pre-synaptic current is described as $I_{syn}=-g_{syn}r(v-V_{syn})$, where V$_A=0$~mV and V$_G=-65$~mV.
$g_{syn}$ is the maximal conductance, $g_A$ for excitatory, and $g_G$ for inhibitory synapses.
$r$ is the gating variable and follows a first-order kinetic dynamics: $\tau_{syn} dr/dt=-r+D\sum_j\delta(t-t_j)$, where $\tau_A=5.26$~ms, $\tau_G=5.60$~ms and the summation over j stands for the neighbor's pre-synaptic spikes at the previous time step $t_j$. $D$ is taken, without loss of generality, equal to $0.05$. Also, all neurons are subject to an independent noisy spike train described by a
Poisson distribution with rate $R$. The input mimics excitatory synapses from neurons that are not included in the populations. For Population $1$ we have employed $R=3000$~Hz and for Population $2$, $R=2400$~Hz with a maximal conductance set to $g^{Poisson}_{A}=0.6$~nS.

For the connectivity, each neuron, excitatory or inhibitory, receives $50$ randomly chosen synapses from other neurons within the same population. Excitatory and inhibitory conductances are set to $g^1_A=3$~nS and $g^1_G=16$~nS for population $1$ and $g^2_A=0.8$~nS and $g^2_G=16.4$~nS for population $2$. For the connectivity between populations, each neuron in population $1(2)$ receives $20$ randomly chosen synapses from excitatory neurons from population $2(1)$. The conductances are set to $g^{12}_A=0.15$~nS and $g^{21}_A=4$~nS, from $1$ to $2$ and from $2$ to $1$, respectively. For the simulations in Fig.~\ref{fig:parameters}, where feedback and feedforward strength connections were decreased or increased by $50\%$, we decreased or increased $g^{12}_A$ and $g^{21}_A$ simultaneously by the same amount.

The model was implemented in a C++ code and simulated using the Euler method, with a time step of $5\times 10^{-2}$~ms. Our code can be found at \url{https://github.com/ldallap/BidirectionalGrangerModel}.

\subsection{Time series analysis in the frequency domain: Power, Coherence and Granger causality}

To determine  functional connectivity  at the population level we analyzed the time series generated from the average membrane potential of excitatory neurons in each population: $V_X=\sum_{i=1}^{400} v_i$, where $v$, the cell's membrane potential, is given by Eq.~\ref{dv/dt}, and $X$ stands for the population $(X= 1,2)$. Power, Coherence, and Granger causality spectral analyses of our simulated time series were calculated using a similar methodology employed in Refs~\cite{Brovelli04,Matias14} and the MVGC Matlab toolbox~\cite{Barnett14}. 
The autoregressive modeling method (MVAR) employed here models the value of a stochastic process at current time $t$ in terms of its $p$ past values at times $t_1$, ... , $t_p$. The regression coefficients represent the predictable structure of the data, whereas the residuals represent the unpredictable structure (see Ref.~\cite{Barnett14} for more details about the Granger causality).

To estimate the spectral analysis from the LFP time series, the MVAR requires the ensemble of single-trial time series to be treated as produced from a zero-mean stochastic process.
Therefore, we have analyzed the simulated time series of each population as if it was generated by $100$ repetitions of $480$~ms each (which is equivalent to $100$ time series of $96$ points with a sample rate of $200$~Hz). It is also necessary to determine an optimal order for the MVAR model. For this purpose, we obtained the minimum of the Akaike Information Criterion (AIC)~\cite{Akaike74} as a function of model order (which we allowed to vary from 1 to 10, i.e. we considered influences up to $50$~ms in the past). 

We calculated the spectral matrix elements $S_{XY}(f)$, with $X=1,2$ and $Y=1,2$ from which the coherence spectrum $C_{12}(f)=|S_{12}|/[S_{11}(f)S_{22}(f)]$ and the phase difference spectrum $\Phi_{lk}(f)=\tan^{-1}[\text{Im}(S_{lk}/\text{Re}(S_{lk}))]$ were calculated. A peak of
$C_{12}(f)$ indicated synchronized oscillatory activity at the peak frequency $f_{peak}$, with a time delay $\tau_{lk}=\phi_{lk}(f_{peak})/(2\pi f_{peak})$.
%\textcolor{red}{[Mauro: what is the difference between $\Delta\Phi$ and $\phi$?]}. 
Directional influence from population $X$ to population $Y$ was assessed via the Granger causality spectrum $GC_{X\rightarrow Y}(f)$. 
We say that if the $p$ past values of $X$ does convey information about the future of $Y$ above and beyond all information contained in the past
of $Y$ then $X$ Granger-causes $Y$. On the other hand, %\textcolor{red}{[is this previous ``if'' correct?]} 
$X$ does not Granger-cause $Y$ if and only if $Y$, conditional on its own past, does not depend on the past of $X$.

Since $GC_{1\rightarrow 2}(f)$ is calculated independent of $GC_{2\rightarrow 1}(f)$,  to quantify the possible asymmetry between the influence from 1 to 2 and from 2 to 1, we can determine, for each frequency, the directed asymmetry index (DAI) between the two areas. 
DAI is computed as the normalized difference between spectral GC influences as in Ref.~\cite{Bastos15}:
\begin{equation}
    DAI_{1\rightarrow 2}(f)= \frac{GC_{1\rightarrow 2}(f)-GC_{2\rightarrow 1}(f)}{GC_{1\rightarrow 2}(f)+GC_{2\rightarrow 1}(f)}.
    \label{eq:DAI}
\end{equation}

\subsection{Neuronal directional information (DI) estimation}

To compute the directional information (DI) from the spiking neuron series we use the same methodology proposed by Tauste Campo et al.~\cite{Campo19}. The method consists of estimating the directional flow of information between a pair of spike trains recorded simultaneously, which are assumed to be generated according to a Markovian process. Given a spike train pair ($X^T,Y^T$) of length $T$, time delay $D\geq0$, and Markovian orders M$_1>0$ and M$_2>0$, the information-theoretic measure ($I(D)$) quantifies the information that the past of $X$ at a delay $D$ has about the present of $Y$, i.e., $I(D)=I(X^{T-D}\rightarrow Y^T)$. The $I(D)$ significance is then determined via nonparametric testing of maximizing-delay statistics which returns a statistic value and the maximizing delay. Finally, DI is defined as the spike-train pairs associated with significance estimators ($\alpha=0.05$). In this context, feedforward, feedback, and bidirectional are defined when $X\rightarrow Y$, $X\leftarrow Y$, and $X\leftrightarrow Y$ are significant, respectively.

DI was estimated over spike-train time-series of $10$~s long from $100$ randomly selected excitatory neurons of each population, i.e., $10000$ spike-train pairs. The spike-train was binarized in $1$~ms bins and the time series was divided into $10$ non-overlapping time bins. DI was performed at time delays $D=0,2,4,...,20$ with maximum memory $M_{1,2}=2$ in accordance with Ref.~\cite{Campo19}. The Matlab code with directional information implementation can be downloaded from \url{https://github.com/AdTau/DI-Inference}.

%%%%%%%%%%%%%%%%%%%%%%%%%%%%%%%%%%%%%%%
%%%%%%%%%%%%%%%%%%%%%%%%%%%%%%%%%%%%%%%
%%%%%%%%%%%%%%%%%%%%%%%%%%%%%%%%%%%%%%%

\begin{figure}[t]%
\includegraphics[width=1.\columnwidth,clip]{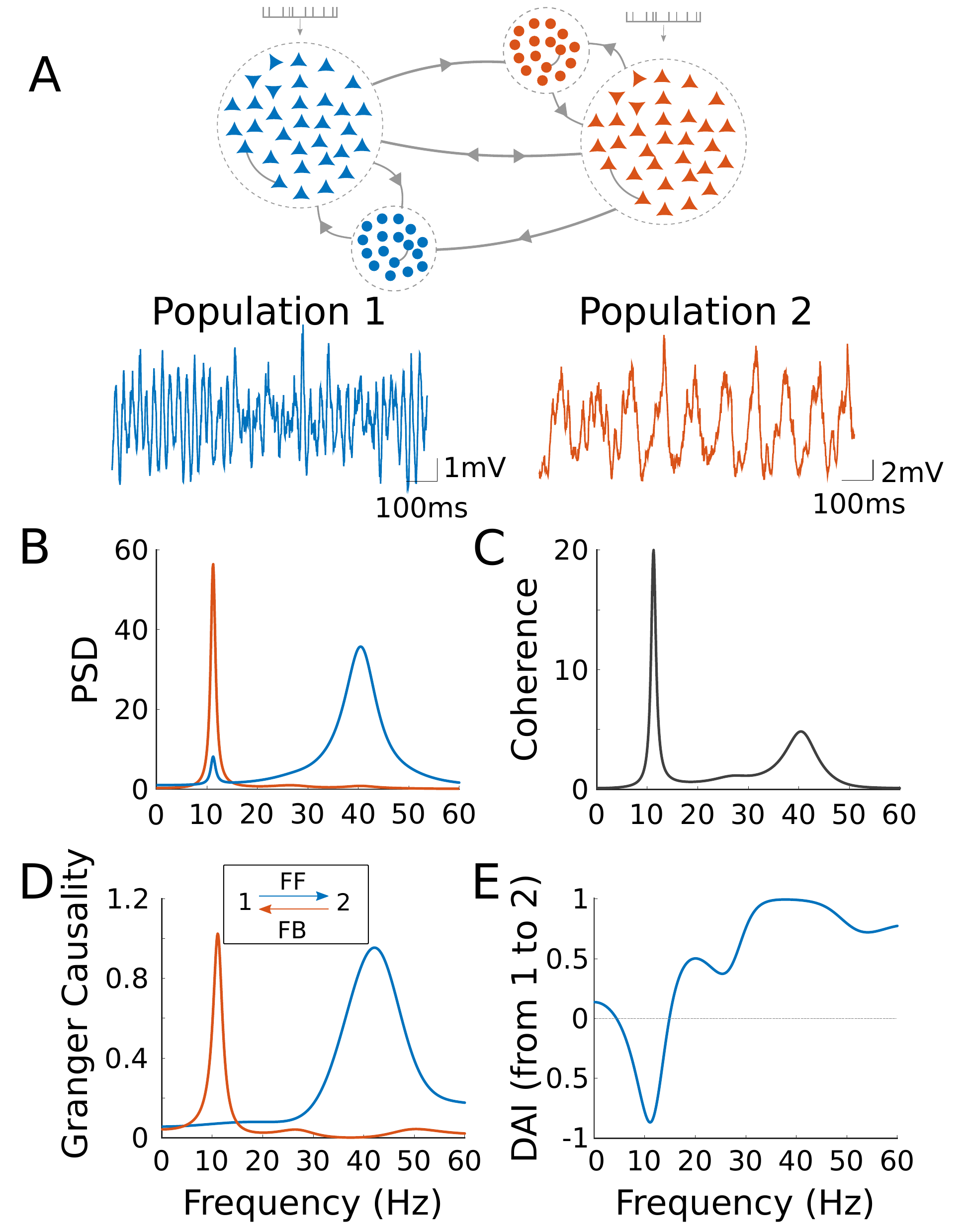}
\caption{\label{fig:coupled} Frequency-specific feedforward and feedback interactions. (A) Schematic representation of two cortical areas coupled in a bidirectional configuration and their average membrane potential.
(B) Power spectral density of each population.
(C) Spectral coherence (equivalent to the cross-spectral density) between the two areas.
(D) Spectral Granger causality in both
directions, showing that each of the peaks found in (C) corresponds to a particular direction of influence.
(E) The directed influence asymmetry index, or DAI profile of the functional connection, which is obtained by normalizing the difference between the
two GC profiles (see Eq.~\ref{eq:DAI}), can be used to characterize a directed functional connection between two cortical areas.}
\end{figure}%

\section{Results}
\label{results}

To study the effect of unidirectional influence in one frequency band and the opposite direction of influence in another frequency band, 
we first simulate two uncoupled  populations that could mimic, without loss of generality, areas V1 and V4 in the visual cortex. 
This would allows us, when we connect the two populations in the following sections, to identify Population 1 as lower-order areas and Population 2 as higher-order areas in the hierarchical organization and reproduce experimental results of feedforward and feedback interactions.

We adjust the parameters of each region to obtain Population 1 oscillating in the gamma band ($\sim 40$~Hz) and Population 2 oscillating in the alpha band ($\sim 10$~Hz) when isolated.
In fact, several visual tasks have been related to an increase in the gamma activity of V1~\cite{Henrie05,Jia13a}.
Since the Granger causality from one network to the other is predominantly around the network ongoing oscillation frequency~\cite{Faes17}, we expect to obtain a causal flow from 1 to 2 at the gamma band and from 2 to 1 at the alpha band when we turn the connections on between the two networks in the following sections.
The important parameters to adjust the rhythms are the internal synaptic conductances, and an external constant current applied to Population 1 (see section~\ref{methods} for more details about the model). 
Fig.~\ref{fig:uncoupled} shows an illustrative example of the two populations, their oscillatory activity represented by the average membrane potential and the power spectra associated with them.

\subsection{Frequency-dependent feedforward and feedback interaction}
\label{coupled}
By connecting the two populations with chemical synapses in such a way that we know the structural connectivity (see Fig.~\ref{fig:coupled}A) we can measure their power spectrum as well as the synchronization and causal relations between their activity to infer the functional connectivity of our motif. In Fig.~\ref{fig:coupled} we show the temporal evolution of the average membrane potential and its correspondent spectral analysis. First, the power spectrum (Fig.~\ref{fig:coupled}B) of each signal indicates that, due to the coupling,
the two populations oscillate both in the gamma and alpha bands. Second, the coherence spectrum (Fig.~\ref{fig:coupled}C), characterizing the cross-correlation in the frequency domain, shows that the activities of the two areas are synchronized predominantly around $10$ and $40$~Hz. Finally, the spectral Granger causality (Fig.~\ref{fig:coupled}D) profiles show that the statistical causal influence is from Population $1$ to Population $2$ ($1 \rightarrow 2$) at the gamma band and the other way around ($2 \rightarrow 1$) at the alpha band. This means that each of the peaks found in the coherence is related to a particular direction of influence.

The time delay $\tau_{lk}$ associated with each frequency peak in the coherence spectrum can be calculated from the phase spectrum (see Methods for more details). We have found that the time delay of the feedforward direction
is $\tau_{12}=3.6$~ms (related to $f_{peak}=40.5$~Hz in Fig.~\ref{fig:coupled}C), whereas the time delay of the feedback interaction is $\tau_{21}=5.3$~ms (related to $f_{peak}=11.3$~Hz in Fig.~\ref{fig:coupled}C).

The directed influence asymmetry index (DAI), which is obtained by normalizing the difference between the
two GC profiles (see Eq.~\ref{eq:DAI}), quantifies the asymmetry between the two directions of influence for each frequency~\cite{Bastos15}. The DAI profile at Fig.~\ref{fig:coupled}E 
corroborates that the directed functional connection between the two cortical areas is predominantly from 1 to 2 at the gamma band ($30-60$~Hz) and from 2 to 1 at the alpha band ($7-13$~Hz).

\begin{figure}[!ht]%
\includegraphics[width=0.9\columnwidth,clip]{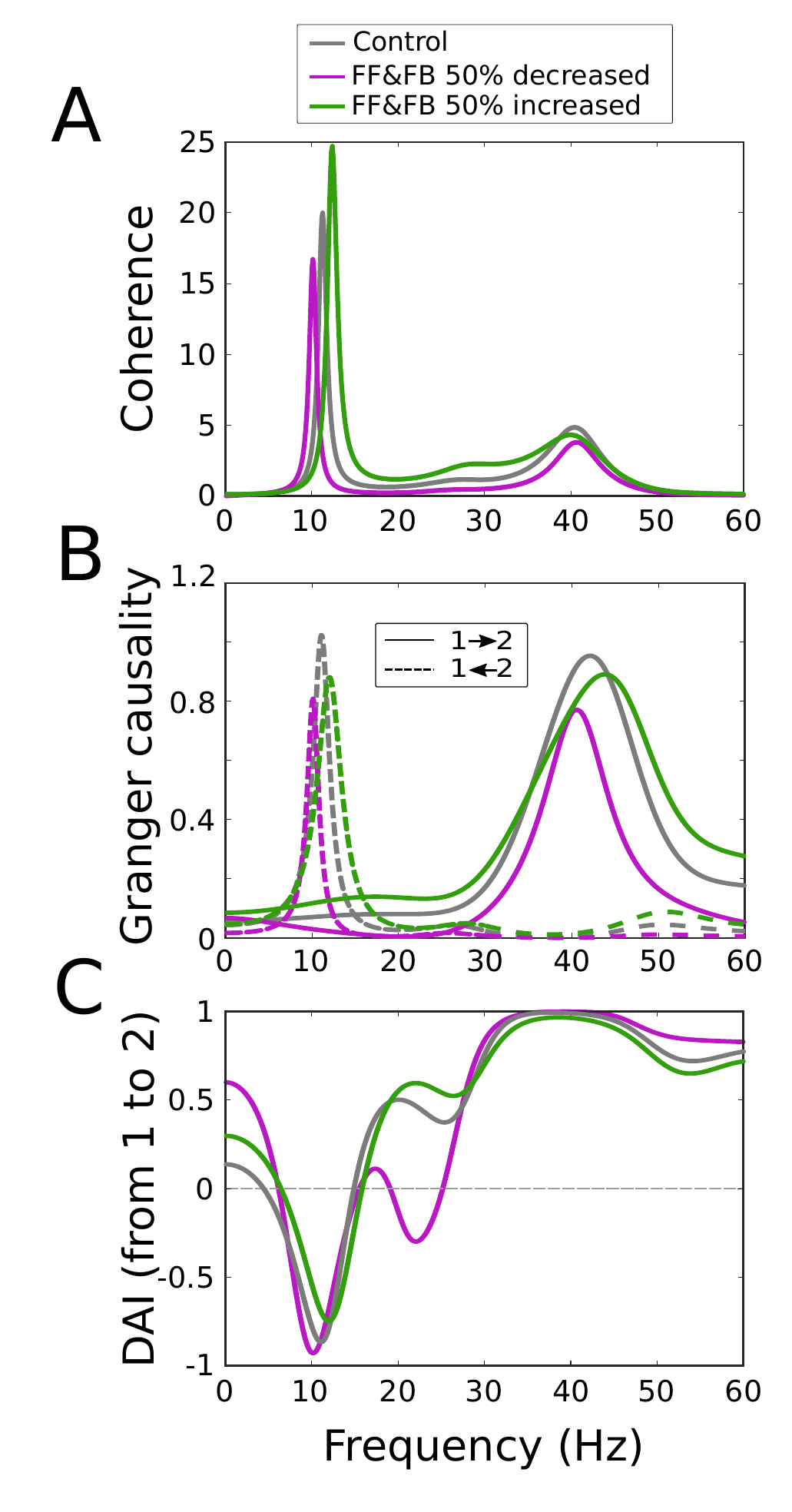}
\caption{\label{fig:parameters} 
Robustness of the bidirectional coupling in the parameter space. (A) Spectral coherence. (B) Spectral Granger causality in both
directions, showing that each of the peaks found in (C) corresponds to a particular direction of influence. (C) The directed influence asymmetry index (DAI) profile of the functional connection.}
\end{figure}

To verify if the results are robust against model parameters, we modified the coupling parameters (between populations) in the simulation and reanalyzed the time series. In Fig.~\ref{fig:parameters} we show that an increase (decrease) in the inter-areal connections slightly shifts the frequency of synchronization and influence but the overall results remain the same.

\subsection{Feedforward and feedback influences at the neuronal scale}
\label{neuronal}

\begin{figure*}[!ht]%
\includegraphics[width=0.8\textwidth,clip]{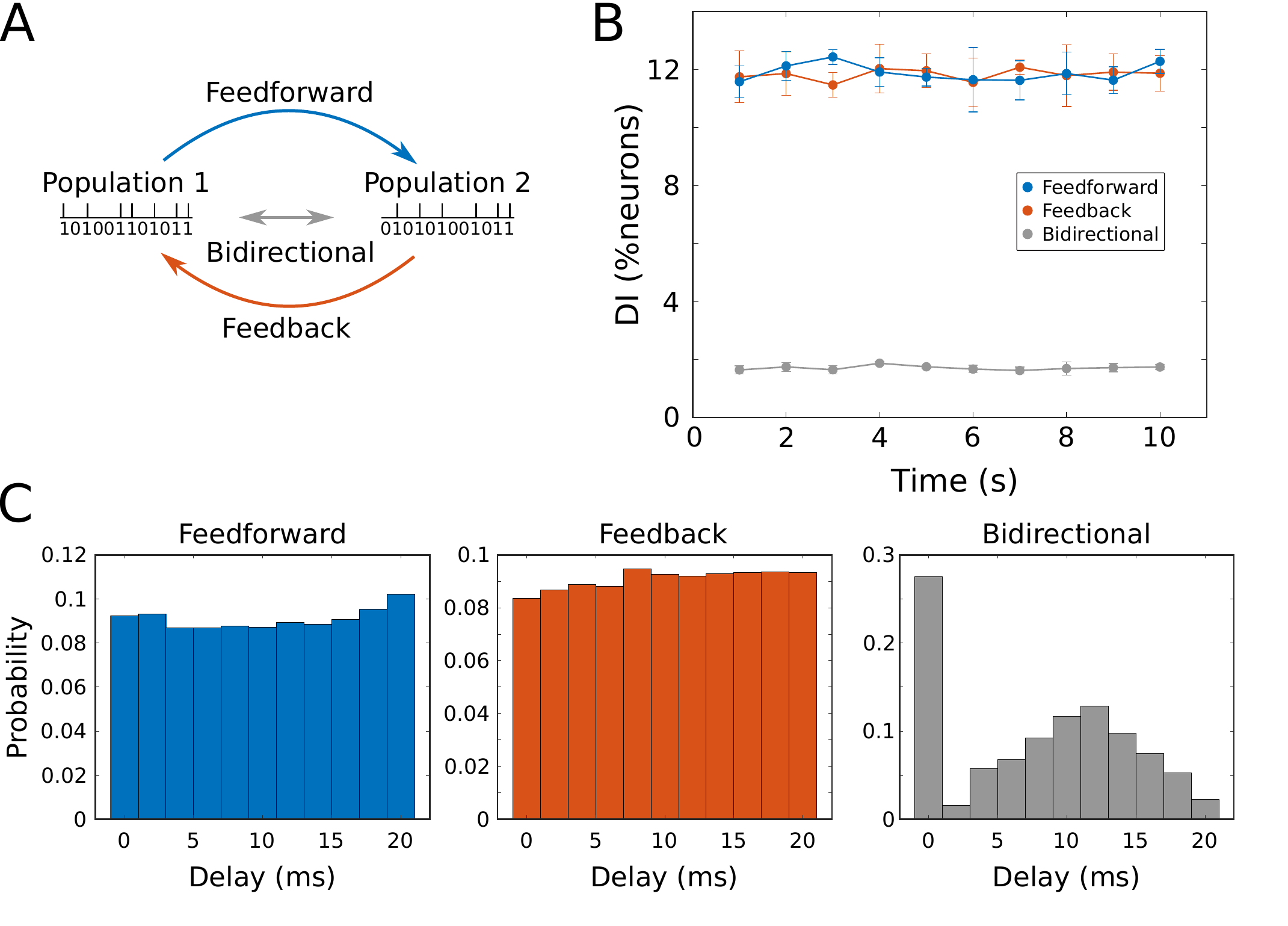}
\caption{\label{fig:DI} Directional information between spike trains pairs. (A) Schematic representation of two spike trains from distinct populations: the directional information (DI) can be via feedforward (blue, Population $1$ $\rightarrow$ Population $2$), feedback (orange, Population $1$ $\leftarrow$ Population $2$) or bidirectional (gray, Population $1$ $\leftrightarrow$ Population $2$). (B) Percentage of neuron pairs involved in feedforward, feedback and bidirectional assessed through DI (see Section~\ref{methods} for details). (C) Delay distribution associated with the feedforward, feedback and bidirectional interactions depicted in (B).}
\end{figure*}

As suggested by Bastos et al.~\cite{Bastos15} 
future experimental studies might test causality influences directly with simultaneous multiarea multilayer recordings of local field potential (LFP) and spikes.
Therefore, modeling the observed phenomena with a spiking neuron network allows us to investigate specific properties of the spiking trains which could be compared in the future with similar experimental data. This simultaneous investigation of LFP and spike would not be possible in a firing rate model.
We address the question of causal relation at the neuronal level
by using a nonparametric directed information-theoretic (DI) measure ~\cite{Campo19,campo2020review}. DI allows the estimation of spiking direct influences from one population to another (Fig.~\ref{fig:DI}A). By analyzing the interaction of $10000$ neuron-pairs ($100$ neurons from Population $1$ with another $100$ neurons from Population $2$, see section~\ref{methods} for details), we can estimate the percentage of neurons from Population $1$ that influences the neurons from Population $2$ (and vice-versa) through feedforward (feedback) interactions. Moreover we can quantify the percentage of these connections that are bidirectional. In our population model, by estimating DI in $1$-s long non-overlapping time windows over $10$-s long time series (see section~\ref{methods} for details), we obtained that both feedforward and feedback interactions are mediated by approximately the same number of neuron-pairs, around $\sim12\%$ (Fig.~\ref{fig:DI}B). For bidirectional communication there are less neurons involved, around $\sim2\%$ (Fig.~\ref{fig:DI}B). 

To measure the communication delay through which neurons interact,
we estimated delayed versions of the directed information-theoretic
measure in both directions at short time delays $D=0, 2, 4, ..., 20$~ms. The probability of finding an interaction for each delay is shown in Fig.~\ref{fig:DI}C. For feedforward and feedback interactions there is no preferred delay of communication (Fig.\ref{fig:DI}C, left and middle), while for the bidirectional case the interaction occurs mostly at zero-lag, followed by a less pronounced peak around $10$ to $12$~ms (Fig.\ref{fig:DI}C right). It is worth mentioning that these communication delays at the neuronal level 
are calculated at the time domain and not at a specific frequency. Therefore, they are different from the time delays $\tau_{lk}$ associated to each frequency band at the populational scale.

%%%%%%%%%%%%%%%%%%%%%%%%%%%%%%%%%%%%%%%
%%%%%%%%%%%%%%%%%%%%%%%%%%%%%%%%%%%%%%%
%%%%%%%%%%%%%%%%%%%%%%%%%%%%%%%%%%%%%%%
\section{Concluding remarks}
\label{conclusions}

To summarize, we have shown that a simple but biophysically plausible model of two bidirectionally connected spiking-neuronal populations can present unidirectional Granger causality in the gamma frequency band and the opposite direction of causal flow in the alpha frequency band.
Our model qualitatively reproduces experimental results verified with
electrocorticogram (ECoG) in macaques and magnetoecephalogram (MEG) in humans of feedforward and feedback influences through distinct frequency bands~\cite{Bastos15,Michalareas16}. In particular, there are two advantages to use spiking-neuron models: i) investigate detailed mechanisms underlying the phenomena such as neuronal variability, and synaptic plasticity within each population; ii) analyze statistical properties of neuronal spiking series which could also be investigated in experimental setups.

Our results are a first step to explore more realistic and detailed spiking-neuron networks including details about the different cortical layers~\cite{Potjans14,Lee15}. It would be possible to use our model to explore a large-scale network with many cortical areas in order to explore hierarchic properties between visual regions, similarly to what has been done by Mejias et al.~\cite{Mejias16} with a firing rate model for each cortical layer.

Furthermore, our findings open new avenues to explore the role of phase diversity on computational properties of brain signals, a subject that has gained the attention of the neuroscience community in the last years ~\cite{Maris13,Maris16}. In light of anticipated synchronization ideas ~\cite{Voss00,Matias14,DallaPorta19} it is possible to explore the mechanisms underlying the phase relation between the two populations in each frequency band both experimentally and numerically.

\begin{acknowledgments}

The authors thank Adri\`a Tauste Campo for insights and data analysis support.
The authors also thank CNPq (grants 432429/2016-6, 301744/2018-1, 425329/2018-6 and 311418/2020-1), CAPES (grants 88881.120309/2016-01, 301744/2018-1 and 425329/2018-6), and FACEPE (grant No.
APQ-0642-1.05/18) for financial support.
LDP acknowledges support from the Spanish Ministry of Science and Innovation through the MICINN under grant BFU2017-85048-R.
This article was produced as part of the activities of FAPESP Center for Neuromathematics (grant 2013/07699-0, S. Paulo Research Foundation).

\end{acknowledgments}

\bibliography{matias}

\end{document}